\documentclass[aps,amsfonts,nofootinbib,superscriptaddress,preprint]{revtex4-1}
\usepackage[normalem]{ulem} 
\makeatletter
\makeatletter
\usepackage{tikz}
\usetikzlibrary{decorations.pathmorphing, shapes.misc, shapes}
\usepackage{amsmath}
\usepackage{tikz-feynman}
\tikzfeynmanset{compat=1.1.0}
\usepackage{tikz-feynman} 
\tikzfeynmanset{compat=1.1.0} 
\usepackage{feynmp-auto}
\usepackage{cleveref}
\usepackage{amssymb}
\usepackage{amsbsy}
\usepackage{bm}
\usepackage{epsfig} 
\usepackage{color}
\usepackage{slashed}
\usepackage{soul}
\usepackage{comment}
\usepackage{auncial}
\makeatletter
\usepackage{appendix}
\usepackage{epsfig} 
\usepackage{ulem}
\usepackage{pgfplots}
\newcommand{\stkout}[1]{\ifmmode\text{\sout{\ensuremath{#1}}}\else\sout{#1}\fi}
\newcommand{\ee}{\end{equation}}
\newcommand{\bb}{\begin{equation}}
\newcommand{\eqb}{\begin{eqnarray}}

\newcommand{\eqf}{\end{eqnarray}}

\DeclareMathOperator{\Tr}{Tr}
\usepackage[T1]{fontenc}
\usepackage{comment}
\usepackage[normalem]{ulem} 
\makeatletter
\usepackage{xcolor}
\usepackage{mdframed}
\usepackage{amsmath}
\usepackage{amssymb}
\usepackage{appendix}
\usepackage{amsbsy}
\usepackage{tikz}
\usepackage{booktabs}
\usetikzlibrary{decorations.pathmorphing, shapes.misc, shapes}
\usepackage{amsmath}
\usepackage{tikz}
\usepackage{feynmp-auto}
\usepackage{cleveref}
\usepackage{bm}
\usepackage{epsfig} 
\usepackage{color}
\usepackage{slashed}
\usepackage{soul}
\usepackage{comment}
\usepackage[normalem]{ulem} 
\makeatletter
\usepackage{amsmath}
\usepackage{amssymb}
\usepackage{appendix}
\usepackage{amsbsy}
\usepackage{bm}
\usepackage{epsfig} 
\usepackage{color}
\usepackage{slashed}
\usepackage{soul}
\begin{document}
\title{Modeling Confinement in Analogy with Black Holes}
\author{J. Gamboa}
\email{jorge.gamboa@usach.cl}
\affiliation{Departamento de F\'{\i}sica, Universidad de Santiago de Chile, Santiago 9170020, Chile}

\begin{abstract}
It is proposed that confinement in QCD be analyzed analogously with the trapping of matter inside a black
hole. To this end, we compare the effective gravitational actions (including quantum corrections)
with the QCD action under the assumption of instanton dominance. The similarities between the
two actions are discussed, and—while the analogy is formal, given the vastly different nature of the
systems—it establishes interesting formulas that may shed light on longstanding problems from a
new perspective.
\end{abstract}
\maketitle
\section{ Introduction}

Field theories in the Carroll regime provide a way to describe the low-energy regime non-perturbatively. Although the contraction $c \to 0$ is a longstanding result in group theory \cite{levy}, only in recent years have possible applications emerged within quantum field theory and gravity. These ideas are deeply intriguing and rich with subtleties, making the study of Carrollian theories challenging and intellectually provocative, particularly regarding their physical applications. 

The low-energy regime of QCD serves as an example of an infrared problem that can be framed within the context of a Carrollian theory—a perspective that is one of the goals of this research. To build an intuitive understanding, consider a hadron with a radius $R$ containing quarks.  QCD is a perturbatively renormalizable and well-defined quantum field theory in the high-energy regime, where $ R \to \infty$. However, when $R \sim 1 \, \mathrm{fm}$, the quarks inside act as point-like particles, and the volume naturally becomes an infrared cutoff.

An example of this statement is the equivalence between the BPST instanton \cite{polyakov} and the scalar field with self-interactions. This example, which we will discuss in the next section, is very illustrative because it contains all the ingredients necessary to understand some non-perturbative effects of QCD. 

While instantons contribute significantly to the QCD vacuum, they cannot fully account for confinement \cite{review1}. Additional arguments are required, which could be borrowed from other fields, such as black hole physics. The interior of a black hole is naturally confined, and it is not far-fetched to suggest that the interior of a hadron may formally resemble the interior of a black hole.

The interior of a black hole is a region causally disconnected from the exterior because the event horizon acts as a potential barrier that prevents matter from escaping the black hole. Similarly, within a hadron, quarks are confined and cannot escape beyond the radius defined by $\Lambda_{QCD}$, which is a ``hadronic horizon''. In a black hole, matter can ``escape'' via the Hawking radiation mechanism, while in a hadron, the analogous phenomenon is deep inelastic scattering, which reveals how ``radiation'' (quarks and gluons) emerges from the interior of a hadron.

A scalar field theory can describe a black hole with temporal derivatives only due to the Kantowski-Sachs geometry \cite{KS, brehme}. Similarly, QCD might be explained analogously using the equivalence between scalar fields and BPST instantons. In other words, this phenomenon still seems to require a proper interpretation, and the emergence of an effective action appears to be inherently topological.

The paper is organized as follows: Section II shows how the BPST instantons are related to instantons coming from a scalar field with \( \varphi^4 \) coupling. We also explain how the coupling constants of pure Yang-Mills theory and scalars connect. Additionally, the role of the infrared cutoff is discussed in detail in this section. In Section III, we motivate the connection between Carrollian theory and QCD in various contexts. Section IV discusses the relationship between the Carroll regime and black holes, and the dynamics of black holes as an effective theory. In Section V we discuss de QCD-Black hole correspondence, Section VI discusses the mapping between the effective actions of QCD and black holes, and a confinement criterion is proposed. In section VI the deconfinement of black holes is explained and finally, in section VII, we summarize our results.

\section{Scalar Field Theory in the Carroll Regime and BPST Instantons}
The following is an initial framework to explore the concept. We consider a scalar field theory in the Carroll limit, where spatial velocities are negligible \footnote{In a pioneering and somewhat overlooked paper, John Klauder introduced this approach in the early 1970s \cite{klauder}.}. In this limit, 
the theory effectively reduces to a one-dimensional system in the temporal dimension, allowing us to investigate possible classical solutions that may resemble 
instantons. By examining the Euclidean action and analyzing configurations that interpolate between different vacua, we aim to identify stable, localized solutions.

In the Carroll limit, where spatial velocities approach zero, we consider a scalar field theory described by the following Lagrangian:

\bb
L = \frac{1}{2} \dot{\phi}^2 + \frac{\lambda}{4} (\phi^2 - a^2)^2. \label{1}
\ee

This theory is defined in a 4D space, but we will show that it can still admit classical solutions that behave as instantons.

\subsection{The Carroll Limit and Dimensional Reduction}
As mentioned above, this Lagrangian resembles a one-dimensional quantum mechanics problem with a potential. with a potential given by:

\bb
V(\phi) = \frac{\lambda}{4} (\phi^2 - a^2)^2, \label{2}
\ee
but with the important technical detail, we are neglecting spacetime's spatial volume.

The potential \( V(\phi) \) has two degenerate vacua at \( \phi = \pm a \), suggesting the possibility of instanton solutions that interpolate between these vacua.
\vskip 0.25cm

To find instanton solutions, we consider the Wick rotation to Euclidean time \( t \rightarrow i \tau \), transforming the Lagrangian into:

\[
L_E = \frac{1}{2} \left( \frac{d\phi}{d\tau} \right)^2 + \frac{\lambda}{4} (\phi^2 - a^2)^2. \label{3}
\]

The Euclidean action is then:

\bb
S_E = \int d\tau \left( \frac{1}{2} \left( \frac{d\phi}{d\tau} \right)^2 + \frac{\lambda}{4} (\phi^2 - a^2)^2 \right). \label{4}
\ee

An instanton solution is a configuration \( \phi(\tau) \) that minimizes this Euclidean action while connecting the vacua at \( \phi = -a \) and \( \phi = +a \).
\vskip 0.25cm

Since \( \phi(\tau) \) interpolates between different vacua, it behaves as an instanton. These solutions have finite Euclidean action \( S_E \), satisfying the requirement for instanton solutions, which typically represent tunneling events between vacua in quantum field theory.

However in the Carroll limit for a scalar field, where spatial derivatives are negligible, a ``weakness'' arises in the theory due to a volume divergence that could nullify the tunneling probability. This occurs because the theory lacks spatial variation terms, introducing an explicit dependence on the system's spatial volume  $V$ \cite{justo}.

However, the field theory in the Carroll limit is special because it comes from birth with the volume divergence that must be regularized (in fact, it is an infrared cutoff).

We want to explain this in a little more detail: in the Carroll limit, the Euclidean action $ S_E$ for a scalar field can be expressed as \cite{coleman}:

\bb
S_E = V\cdot S_{\text{local}}, \label{5}
\ee
where \( V \) is the spatial volume of the system, and  $S_{\text{local}}$ is the local contribution to the Euclidean action, depending only on the temporal variable. 

The tunneling probability \( P \) is typically calculated as:

\bb
P \sim e^{-V\cdot S_E}. \label{6}
\ee

If $V$ is considered infinite, then the tunneling probability becomes zero:

\bb
P \sim e^{-V \cdot S_{\text{local}}} \rightarrow 0 \quad \text{as} \quad V \rightarrow \infty. \label{7}
\ee

This result implies that, in the absence of a finite spatial limit, tunneling would be impossible.

A key point is that in the Carroll limit—where spatial derivatives are eliminated—we are effectively in a strong coupling regime where spatial degrees of freedom are confined to infrared effects. This infrared cutoff, which emerges in the Carroll limit, acts as a minimal energy scale, and its presence implies that the theory can no longer be fully analyzed using conventional perturbative methods.

\subsection{Role of the Infrared Cutoff}

To prevent this volume divergence from suppressing the tunneling probability, we introduce an infrared \emph{cutoff}, interpreting \( V \) as an effective volume \( V_{\text{IR}} \). This cutoff limits the maximum spatial scale and regularizes the theory in the low-energy limit.
Thus, the Euclidean action becomes:

\bb
S_E = V_{\text{IR}} \cdot S_{\text{local}}, \label{8}
\ee
allowing a finite tunneling probability:

\bb
P \sim e^{-V_{\text{IR}} \cdot S_{\text{local}}}. \label{9}
\ee

The volume divergence in the Carroll limit represents a ``weakness'' that would lead to a zero tunneling probability without an infrared cutoff. By regularizing the spatial volume with a cutoff $V_{\text{IR}}$, we obtain a finite tunneling probability and a physically coherent system description in the Carroll limit.

\subsection{``Relating'' Scalar Fields and $SU(2)$ Yang-Mills Theory}
\maketitle

This subsection will give direct and simple arguments to show the mapping between scalar and $SU(2)$ self-dual Yang-Mills theory \cite{justo,calvo}. 

First, let us  write the Lagrangian density (\ref{2}) 
in terms of dimensionless variables by the rescaling 
\bb
{\bar \varphi} \to \frac{ \sqrt{\lambda}}{\mu} {\bar \varphi}, ~~~~~t \to \mu~ t , \label{10}
\ee 
 and the action becomes
\bb
S_S = \frac{1}{\lambda} \int d{\bar t} \left[\frac{1}{2} {\dot {\bar \varphi}}^2+ \frac{1}{4} \left({\bar \varphi}^2 - 1 \right)^2 \right]. \label{11}
\ee

Now, we could compare this action with the Yang-Mills one, taking into account the following:
\begin{itemize} 
\item Since the Yang-Mills description we are looking must contain instanton solutions and the self-duality condition
\bb
F_{\mu \nu} = {\tilde F}_{\mu \nu} \label{selfdua}
\ee
must be satisfied.
\item  If we can find a potential that satisfies (\ref{selfdua}), then we not only solve the condition (\ref{selfdua}), but we automatically have a solution of the Yang-Mills equations. Since in the Euclidean space $SO(4) \simeq SU(2) \times SU(2)$, the Yang-Mills potentials are a possible representation of $ (\frac{1}{2}, \frac{ 1}{2}) $ of $SU (2) \times SU(2)$. 

The tensor $F_{\mu \nu}$ belongs to a reducible representation $(1,0) \oplus (0,1)$ so that the quantities can represent the conditions of self and anti-self dualities 
\eqb 
F^a_L &=& \eta^a_{\mu \nu} \left( F_{\mu \nu} + {\tilde F}_{\mu \nu}\right), 
\\
F^a_R &=& {\bar \eta}^a_{\mu \nu} \left( F_{\mu \nu} -{\tilde F}_{\mu \nu}\right), 
\eqf
where $\eta$ and ${\bar \eta}$ are the `t Hooft symbols for $SU(2)$  which have the following properties: 
\eqb 
\eta_{a \mu \nu} &=&  \frac{1}{2} \epsilon_{\mu \nu \rho \sigma} \eta_{a \rho \sigma}, \nonumber
\\ 
{\bar \eta}_{a \mu \nu} &=& -\frac{1}{2} \epsilon_{\mu \nu \rho \sigma} {\bar \eta}_{a \rho \sigma} \nonumber 
\\
{\bar \eta}^a_{\mu \nu} (x)  &=&\epsilon^a_{\mu \nu 4} -\delta^a_\mu \delta_{\nu 4}+ \delta^a_\nu \delta_{\mu 4}, \nonumber
\\
\epsilon_{abc} \eta_{b \mu \nu}  \eta_{c \rho \sigma}&=& \delta_{\mu \rho} + \delta_{\nu \sigma} \eta_{\nu \sigma} \eta_{a \nu \rho} -\delta_{\nu \rho} 
\eta_{a \mu \sigma}, \nonumber
\eqf
\end{itemize} 
among other \cite{thooft1}.

Condition (\ref{selfdua}) is a well-known result, but finding a potential that satisfies (\ref{selfdua}) is a highly non-trivial problem that was solved probably by Diakonov \cite{diakonov}. 

The potential is
\bb
A^a_\mu = {\bar \eta}^a_{\mu \nu} x_\nu \frac{[1+ \phi (x^2)]}{x^2+\rho^2}, \label{pot}
\ee 
 where $\phi$ is a scalar field and where $\rho$ is a length scale and later is chosen as $\Lambda_{\text{QCD}}^{-1}$, with $\Lambda_{\text{QCD}} \approx 200 \, \text{MeV}$ being the QCD scale.

The first step to justify Diakonov's ansatz is to write the $SU(2)$ Yang-Mills potential as 
\bb
A^a_\mu = \eta^a_\mu (x) g(x^2). \label{xx1}
\ee
where $ \eta^a_\mu (x)$ is a function that we will determine below using the t'Hooft symbols and $g(x^2)$ is a function to be determined.

The function $\eta^a_\mu$ is obtained as follows: firstly we take the t'Hooft symbols definition
\[
{\bar \eta}^a_{\mu \nu} (x)  = \epsilon^a_{\mu \nu 4} -\delta^a_\mu \delta_{\nu 4}+ \delta^a_\nu \delta_{\mu 4}
\]
and multiplying by $x_\nu$ we obtain 
\bb
\eta^a_\mu (x) = x^a \delta_{\mu 4} - x^4 \delta_{\mu 4} + \epsilon^{a jk} \delta_{\mu j} x_k. \label{xx3}
\ee

By using (\ref{xx1}) and (\ref{xx3}) the tensor $F^a_{\mu \nu}$ becomes 
\bb
F^a_{\mu \nu} = g(x^2)  \left( \partial_\mu \eta^a_\nu -\partial_\nu \eta^a_\mu \right) + 2 g'(x^2) \left( x_\mu \eta^a_\nu - x_\nu \eta^a_\mu\right) + g(x^2) \epsilon^{abc} \eta^b_\mu \eta^c_\nu. \label{xx4}
\ee

Here $g' (x^2) = \frac{d}{dx^2} g(x^2)$ and the Yang-Mills equations become 
\eqb
\partial_\mu F^a_{\mu \nu} (x^2) &+& \epsilon^{abc} A^b _\mu F^c_{\mu \nu} =  \nonumber
\\
&=&\left( 4 x^2 g''(x^2) + 12 g'(x^2) - 2 x^2 g^3 (x^2) \right) \eta^a_\nu + 3g^2 (x^2) \epsilon^{abc} \eta^b_\mu \partial_\mu \eta^c_\nu =0. \nonumber 
\\ 
\label{xx5}
\eqf

If we choose all the components of the group differently, we obtain the following non-linear differential  equation
\bb
2 x^2 g'' + 6 g' +3g^2 -x^2 g^3=0. \label{xx6}
\ee

In order to solve this eq. let's do the following variable change
\eqb
g(x^2) &=& \frac{h(x^2)}{x^2}   \label{xx7}
\\
x^2 &=& \rho^2 e^t, \label{xx8}
\eqf
where $\rho^2$ is a scale parameter introduced in (\ref{pot}).

Replacing (\ref{xx7}) and (\ref{xx8}) we obtain de following equation
\bb
2 {\ddot h} (t)- 2 h(t) -3h^2(t) - h^3(t)=0.
\ee

These equations can be obtained from the following Lagrangian
\eqb
L &=& {\dot h}^2 + h^2 - h^3 + \frac{1}{4} h^4 \nonumber
\\
&=& {\dot h}^2 + \frac{1}{4} h^2 (h-2)^2,  \label{9}
\eqf
and making the shift
\bb
h \to \varphi +1, \label{xx10}
\ee
we obtain
\bb
L = {\dot \varphi}^2 + \frac{1}{4} (\varphi^2 -1)^2. \label{ins001}
\ee

Replacing (\ref{xx7}) and (\ref{xx10}) in (\ref{pot}), we obtain de Diakonov's ansatz, but we also obtain that $\varphi$ is a solution of the instantons in $2D$ equations.

Now we replacing (\ref{pot}) in the Yang-Mills action
$$
S_{YM} = -\frac{1}{4g^2} \int d^4x ~\Tr (F_{\mu \nu}^2),
$$ 
we find 
\bb
S_{YM} = \frac{12 \pi^2}{g^2} \int d\tau \left[ \frac{1}{2}\left(\frac{d\phi}{d\tau}\right)^2 + \frac{1}{4} (\phi^2 -1)^2\right]. \label{ac2}
\ee
where $\tau = \ln \frac{x^2}{\rho^2}$. 

Comparing (\ref{2}) and  (\ref{ac2}) we see that both actions are equivalent if
\bb
\lambda= \frac{g^2}{12\pi^2}, 
\ee
We note that even though \(\rho\) does not explicitly appear in the final value of the pure instanton action---since the size of the instanton does not affect its topological contribution---this parameter controls the action distribution in space.

We finish this section by noting that the $ x ^ 2 $ dependence of the gauge potential corresponds to a set of solutions to the Yang-Mills equations that maps 
exactly to the infrared limit.

From this discussion, we can conclude that:
\begin{itemize}
\item These scalar instantons are the same ones appearing in Belavin et al.'s solution \cite{polyakov}. 
\item This idea suggests that one could study some nonperturbative properties of QCD using only scalar fields.
\item Of course, this is not real QCD where $N=3$, but this model still captures essential aspects that we will explain below from a different perspective.
\end{itemize}

\section{Motivating Carroll Field Theory and QCD}

The Carroll limit can be relevant in Quantum Chromodynamics (QCD) due to its unique spacetime structure, where time decouples from spatial dimensions. This results in a theory where particles do not move spatially, which has interesting implications for confinement phenomena and the structure of strong interactions. 

\noindent Below are some specific reasons:
\begin{itemize}
\item {\it Confinement and Dynamics in Extreme Limits}. In the Carroll regime, particles do not propagate in space, which can simulate the idea that quarks are ``confined'' within a tiny region and cannot be separated or move freely. This resonates with the confinement property in QCD, where quarks are confined within hadrons and cannot exist as free particles at large distances. Thus, the Carroll limit provides an effective model to study confinement as a natural consequence of the limitation on spatial movement.
\item {\it Analogy with Low-Energy States and QCD in Dense Media}. In low-energy QCD systems or dense media (for example, in neutron stars or conditions of dense nuclear matter), the mobility of quarks may be severely restricted. In these conditions, the Carroll limit could represent the behavior of these systems, where the spatial dynamics of the constituents are limited or even ``frozen'' compared to the internal interaction. This can provide a useful approximation for analyzing interactions in QCD without significant spatial propagation.
\item  {\it Mathematical Simplicity for Effective QCD Models}. The reduction of spatial dynamics in the Carroll limit simplifies the theory. It allows the modeling of strongly coupled interactions, such as those in QCD, in a framework where calculations become more manageable. This is especially useful in analyzing effective theories for confined hadrons and gluons, as the Carroll limit reduces the degrees of freedom and allows a controlled study of the theory's behavior.
\item {\it Connection with Confinement Theory and Solitons}. In the Carroll limit, stable solitonic solutions can arise, which is also characteristic of certain solutions in QCD related to confinement and mass generation. In particular, in the interior of a black hole (if interpreted as a Carroll-like geometry), confinement may emerge naturally, and some arguments suggest that the dynamics of gluons and quarks can be modeled through scalar fields and solitons.
\item {\it  Interpretation in Terms of Gauge Variables and Yang-Mills Theory}. The Carroll limit could help simplify the description of gauge theories in QCD, where gluons (the mediators of the strong interaction) can be described in a regime that favors local interactions without propagation, emulating quark confinement. This can be particularly useful in analyzing self-dual Yang-Mills solutions and in studying field configurations related to the QCD vacuum. 
\end{itemize}

\section{Black Holes as Carroll Field Theory}

The AdS-CFT correspondence \cite{maldacena} has indeed highlighted the relevance of black holes in the holographic context, usually associating their horizon and thermal behavior with the gauge theory on the AdS boundary. 

However, the direct exploration of the interior of black holes as a possible “route” to new aspects of physics or field structure has been limited. 
Exploring the black hole interior beyond the framework of classical general relativity, for instance, could open new perspectives on the quantum structure of spacetime, confinement, or soliton formation under extreme conditions—concepts less developed in the usual holographic framework.
\vskip 0.25cm

        A primary motivation for exploring the interior of black holes is its potential to reveal new aspects of physics that remain largely unexplored. While AdS-CFT correspondence and other holographic approaches have successfully illuminated the relevance of black holes, they have primarily focused on the horizon and surrounding regions where dualities are most applicable. In contrast, the {interior}---with its unique conditions, such as extreme curvature, causal disconnection, and potential confinement---remains less understood and rarely addressed within the holographic framework.
        
\vskip 0.25cm

By directly investigating the black hole interior, we could uncover insights into fundamental physics topics like confinement, soliton formation, and even quantum aspects of spacetime structure. These areas could yield perspectives that are challenging to capture solely through horizon-based or boundary interpretations. This exploration of the black hole interior opens a promising path to address unanswered questions and extend our understanding beyond the established limits of current duality-based approaches.

This idea suggests that, in certain extreme limits, scalar fields can effectively capture key aspects of the complex dynamics of Yang-Mills (YM) theory, thereby simplifying the study of hadronic internal structure. Considering a scalar field in the Carroll limit makes it possible to approximate quarks' confined nature and lack of free propagation, providing an alternative and potentially insightful perspective on the strong interactions within hadrons. In this way, exploring confinement and the internal dynamics of gluons and quarks using scalar fields in the Carroll regime allows for a more accessible theoretical treatment under extreme conditions, effectively capturing the essence of these interactions without losing their fundamental properties.
\subsection{Black Holes and Carroll Theory}
In this subsection, we will investigate the dynamics of black holes from the perspective of Carroll's field theory.
Our approach is not the usual one, because we will formulate something that incorporates the intricacies of curvature but has been written to resemble a standard problem in quantum field theory in the Carrollian regime.

Suppose a black hole is described by the Kantowski-Sachs metric, which in natural units ($c = G = 1$) takes the form \\cite{vile}:
\bb
ds^2 = -N^2(t) dt^2 + a(t)^2 dr^2 + b(t)^2 \left( d\theta^2 + \sin^2\theta \, d\phi^2 \right), \label{ks01}
\ee 
where $N(t)$ is the lapse function, 
 $a(t)$ is the scale factor corresponding to the radial direction and $b(t)$ is the scale factor for the angular directions $(\theta, \phi)$.

This metric describes a highly symmetric spacetime, where the radial and angular components evolve independently over time. It is often used to model the interior of a black hole, where the roles of time and space coordinates can interchange due to the causal structure of the spacetime.

This point is worth emphasizing because the Kantowski-Sachs metric, apart from its cosmological applications, can also be understood as an interior solution of a black hole where the symmetry $r \leftrightarrow t$ is satisfied ({\it i.e.}, the equations of motion evolve in $r$ rather than $t$). This observation will become important later on.

The Einstein-Hilbert action is given by:
\bb
S = \int d^4x \sqrt{-g} \left( R - 2 \Lambda \right),
\ee
and substituting the Kantowski-Sachs metric (\ref{ks01}), we obtain the Lagrangian \cite{vile}

\eqb
L&=& \frac{1}{N} \left( 2 b \dot{b}a  \dot{a} + \dot{b}^2 a^2 \right) + N \left( \Lambda b^2 - 1 \right) \nonumber
\\
&=&  \frac{{\dot c}~{\dot b}}{N}  + N \left( \Lambda b^2 - 1 \right), \label{lag1}
\eqf
where  $c=a^2 b$.

With this in mind, let us consider the following change of variables
\bb
c = \frac{1}{\sqrt{2}}(\varphi_1 + \varphi_2), ~~~~~~~~~~~~b= \frac{1}{\sqrt{2}} (\varphi_1 -\varphi_2), \label{change1}
\ee
then the Lagrangian is
\begin{equation}
\mathcal{L} = - \frac{1}{N} \left( {{\dot \varphi}_1}^2 - {{\dot \varphi}_2}^2 \right) + N \left( \Lambda (\varphi_1 - \varphi_2)^2 - 1 \right). \label{333}
\end{equation}

Returning to a new (and simple) change of variables to diagonalize both the kinetic and potential energy terms simultaneously (renaming variables).
\bb
\psi_1 = \frac{1}{\sqrt{2}}(\varphi_1 +\varphi_2), ~~~~~~~\psi_2 = \frac{1}{\sqrt{2}}(\varphi_1 -\varphi_2),
\ee
the Lagrangian becomes
\begin{equation}
\mathcal{L}_0 = - \frac{1}{2N} \left({ {\dot \psi_1}}^{2} - { {\dot \psi_2}}^{2} \right) + N \left( 2 \Lambda {\psi}_2^2 - 1 \right). \label{bh17}
\end{equation}

In this way, the Lagrangian describing the interior of a black hole can be written as a Carrollian theory with two scalar fields. One of them ($\psi_2$) is perfectly well-defined (if we choose $\Lambda < 0$ ), while the second field has negative kinetic energy ($\psi_1$), indicating an instability that we must understand.

Writing the Lagrangian (\ref{bh17}) represents a significant notational simplification. It avoids the technical complications associated with high curvature and facilitates the incorporation of interactions, as in conventional quantum field theory.

A black hole is a mathematical idealization because it only makes sense when it interacts gravitationally with matter (as other objects do, as they do not exist in isolation without interacting with others). For this reason, let us assume that the fields ($\psi_1$) and ($\psi_2$) interact with matter, collectively denoted by $\chi$, through:
\bb
{\cal L}_1 = \frac{1}{2N}{\dot \chi}^2 - \frac{1}{2N} M^2 \chi^2 +\frac{g_1}{2} N \chi^2 \psi_1^2 + \frac{g_2}{2} N \chi^2 \psi_2^2. \label{inter}
\ee
 
 In the gauge $N=1$, the equations of motion are as follows:
 \eqb 
 &&{\ddot \psi}_1 =  - g_1 \chi^2 \psi_1, 
 \\
 && {\ddot \psi}_2 = 4 \Lambda \psi_2+ g_2 \chi^2 \psi_2
 \\
 &&
 {\ddot \chi}= -M^2 \chi + (g_1 \psi_1 + g_2 \psi_2)\chi,
 \eqf 
 besides the constraint
 \bb
 \frac{{\dot \psi_1}^2}{2}-\frac{{\dot \psi_2}^2}{2} -\frac{{\dot \chi}^2}{2} -\frac{M^2}{2}\chi^2 +\frac{g_1}{2} N \chi^2 \psi_1^2 + \frac{g_2}{2} N \chi^2 \psi_2^2=0.
 \ee

This is a set of nonlinear differential equations that can be solved numerically but are written here merely to highlight that, even when it is assumed that the parametrization of the matter is simple, the equations of motion can lead to very complicated solutions. 

\subsection{Quantum Theory of Black Holes as Effective Theory}

In this section, we explore the concept of describing black holes within the framework of Carrollian theory, focusing on our treatment as an effective theory.

The partition function in the proper-time gauge is:
 \eqb
Z &=& \int_0^\infty d N(0) \int \mathcal{D} {\psi}_1 \mathcal{D} {\psi}_2\mathcal{D} \chi\, e^{i S[{\psi}_1, {\psi}_2, \chi,N]} \nonumber
\\
&=& \int_0^\infty d N(0) \int \mathcal{D} {\psi}_1 \mathcal{D} {\psi}_2  
\,{\det\left(1+ \frac{g_1 \psi_1^2 + g_2 \psi_2^2}{- \frac{\partial_r^2}{N^2(0)} +M^2}\right)}^{-\frac{1}{2}}\,\times~ e^{i S[{\psi}_1, {\psi}_2,N]}. \label{bhh1}
\eqf
where in the last line we have integrated out the matter field $\chi$.

Using the identity $\det(1+M)=e^{\Tr\ln(1+M)}=e^{M-\frac{M^2}{2}+ \frac{M^3}{3}+\cdots}$, the perturbative calculation of the effective action gives\footnote{For notational simplicity, we will set $N=1$ and restore it when necessary.}
\eqb
&&\Tr\ln \det(1+M)^{-1/2}=\Tr\left[-\frac{1}{2}G(0)\left(g_1 \psi_1^2+ g_2 \psi_2^2\right)\right] +  \Tr \left[\frac{1}{4}G^2(0){\left(g_1 \psi_1^2+ g_2 \psi_2^2\right)}^2\right]+ \cdots \nonumber \\
&=& 
 \int dt \, \left( -\frac{1}{2} G(0)(g_1\psi_1^2 + g_2\psi_2^2) + \frac{\gamma}{4}( g_1^2 \psi_1^4 +g_2^2 \psi_2^4)+ \gamma \frac{g_1 g_2}{2} \psi_1^2 G(0)^2\psi_2^2 \right) + \cdots \nonumber
\\
 \hspace{0.5cm} 
 &=&\raisebox{1.4cm}{} 
 \raisebox{-1.4cm}{
 \begin{tikzpicture}
    \begin{feynman}
      \vertex (a) at (0, 0);
      \vertex [below left=1cm] (i1) {\(\psi_1\)};
      \vertex [below right=1cm] (i2) {\(\psi_1\)};
      \vertex [above=1cm of a] (b);
      
      \diagram* {
        (i1) -- [plain] (a) -- [plain] (i2),
        (a) -- [plain, dashed, half left, looseness=1.5] (b) -- [plain, dashed, half left, looseness=1.5] (a)
      };
    \end{feynman}
  \end{tikzpicture}
  \hspace{-0.2cm} 
  \raisebox{1.4cm}{$+$} 
  \hspace{-0.2cm}
  \begin{tikzpicture}
    \begin{feynman}
      \vertex (c) at (0, 0);
      \vertex [below left=1cm, xshift=-0.1cm] (j1) {$\psi_1$};
      \vertex [below left=1cm , xshift=0.6cm] (j2) {$\psi_1$};
      \vertex [below right=1cm, xshift=0.1cm] (j3) {$ \psi_1$};
      \vertex [below right=1cm, xshift=-0.6cm] (j4) {$\psi_1$};
      \vertex [above=1cm] (d);
      
      \diagram* {
        (j1) -- [plain] (c),
        (j2) -- [plain] (c),
        (j3) -- [plain] (c),
        (j4) -- [plain] (c),
        (c) -- [plain, dashed, half left, looseness=1.5] (d) -- [plain, dashed, half left, looseness=1.5] (c)
      };
    \end{feynman}
  \end{tikzpicture}
  }
   \hspace{-0.1cm} 
  \raisebox{0cm}{$+\dots$} 
  \hspace{-0.cm}
\eqf
where $\gamma$  is fixed below.

To write the precise form of $G(0)$, we must take into account that in the Carroll limit, the volume element $d^3p$  factorizes and can be written 
as $E_P^3$, where $E_P$ is cutoff. 
In terms of this cutoff,  
\bb
G(0)=  \frac{E_P^3}{2M}, 
\ee
where $E_P^3$ is a volume energy cutoff from $d^3p$ and --in the Carroll limit-- enters as a constant volume. This is an interesting issue because $E_P$ is an infrared cutoff that naturally emerges in a field theory in the Carroll limit. 

However, the infrared cutoff requires the introduction of another ingredient of consistency as we will see below. If we denote this quantity by $\gamma=\Omega^{-1}$, then the effective Lagrangian is
\eqb 
{\cal L}_{\text{eff,int}} &=& -\frac{1}{2} G(0) \left(g_1 \psi_1^2 + g_2  \psi_2^2\right) + \frac{\gamma^4}{4} G^2(0) \left(g_1 \psi_1^2 + g_2  \psi_2^2\right)^2 \nonumber 
\\ &-&
\frac{\gamma^8}{8} G^3(0) \left(g_1 \psi_1^2 + g_2  \psi_2^2\right)^3+ \cdots,
\eqf
and now we identify 
\bb
\gamma= \Omega^{-1}=E^{-1}_P. \label{ident} 
\ee 

The first two bubbles correspond to massive excitations taking the fixed values 
\bb
m_1^2 = g_1 G(0)=g_1\frac{E_P^3}{2 M}, ~~~~~m_2^2 = g_2 G(0)=g_2\frac{E_P^3}{2 M}, \label{defmass} 
\ee
while the dimensionless coupling constants for the quartic diagrams are
\bb
\lambda_1^2 = g_1^2 \frac{G^2(0)}{\Omega^4}=\frac{g^2_1}{4} \left(\frac{E_P}{M}\right)^2, ~~~~~~~~~~\lambda_2^2 = g^2_2 \frac{G^2(0)}{\Omega^4}=\frac{g^2_2}{4} \left(\frac{E_P}{M}\right)^2. \label{lamb12}
\ee

Note that equations (\ref{defmass}) and (\ref{lamb12}) correspond to \uline{the identification of $M$ as the ultraviolet}  cutoff (UV) of the effective theory whose Lagrangian becomes
\eqb
{\cal L}_{eff} &=&-\frac{1}{2}\left({\dot \psi_1}^2 -{\dot \psi_2}^2\right) + (2 \Lambda +m_2^2)\psi_2^2 - \frac{1}{2} m_1^2 \psi_1^2  \nonumber
\\ 
&+&\frac{1}{4}\lambda_1 \psi_1^4 +
\frac{1}{4} \lambda_2 \psi_2^4 + \frac{1}{2} g_1 g_2 \left(\frac{E_p}{2 M}\right)^2 \psi_1^2 \psi_2^2. \label{fin}
\eqf

As an incidental non-trivial note, let us also observe that the Carroll limit implies that, in the absence of spatial propagation, Green's function is necessarily 
$G(0)$, which is consistent with the fact that $\frac{E_P}{M}\ll 1$. 

Thus, the mass $M$ of the scalar $\chi$ effectively plays the role of the UV cutoff in the theory.

Up to this point, our arguments have not addressed the kinetic terms, one of which is negative, nor the mass terms, where one has a ``wrong'' sign \footnote{In principle, there is no ``wrong'' mass sign because the cosmological constant can perfectly be negative.}. However, this can be resolved successfully in either case. 

On the other hand, the stability of the effective theory also arises from the mixed contribution to the effective Lagrangian
\eqb
\kappa \psi_1^2 \psi_2^2=
\raisebox{1.4cm}{} 
\raisebox{-1.4cm}{
\begin{tikzpicture}
\begin{feynman}
    \vertex (c) at (0, 0);
    \vertex [below left=1cm, xshift=-0.1cm] (j1) {$\psi_1$};
    \vertex [below left=1cm , xshift=0.6cm] (j2) {$\psi_1$};
    \vertex [below right=1cm, xshift=0.1cm] (j3) {$\psi_2$};
    \vertex [below right=1cm, xshift=-0.6cm] (j4) {$\psi_2$};
    \vertex [above=1cm] (d);
      
    \diagram* {
    (j1) -- [plain] (c),
    (j2) -- [plain] (c),
    (j3) -- [gluon] (c),
    (j4) -- [gluon] (c),
    (c) -- [plain, dashed, half left, looseness=1.5] (d) -- [plain, dashed, half left, looseness=1.5] (c)
      };
\end{feynman}
\end{tikzpicture}
}
\eqf

We could calculate it by integrating over $\psi_1$, but this presents the difficulty that the kinetic energy for $\psi_1$  is negative. To address this issue, we proceed as follows: since energy is conserved, in the limit where the kinetic energy is minimal, the total energy consists entirely of potential energy\footnote{This could correspond to the limit of complete isotropy, but I have not investigated this fact so far.}. Therefore, we can neglect the kinetic energy of the mode $\psi_1$, treating it as an auxiliary field.

The auxiliary field can be solved algebraically, and we obtain the following two solutions:
\eqb 
 \psi_1&=& 0, \label{cond1}
\\ 
  \psi_1^2 &=& -\frac{1}{\lambda_1}\left[ m_1^2 + \left(\frac{E_P}{2 M}\right)^2 g_1 g_2 \psi_2^2 \right].  \label{cond2} 
  \eqf

We will analyze these conditions below.

\subsection{Solution $\psi_1=0$}
In this case, the effective Lagrangian is\footnote{Here $\cdots$ are contributions independents of $\psi_2$.}
\bb
L_{\text{eff}} =  \frac{1}{2} {\dot \psi_2}^2 + \left( 2 \Lambda +m_2^2 \right) \psi_2^2 + \frac{\lambda^2_2}{4}   \psi_2^4 +\cdots. \label{eff2}
\ee

Thus, the solution (\ref{cond1}) implies an interesting physical result. Let us choose the cosmological constant to be negative, so (\ref{eff2}) becomes:
\bb
{\cal L}_{eff} = \frac{1}{2} {\dot \psi_2}^2 - \kappa^2 \psi_2^2 + \frac{1}{4} \lambda^2_2 \psi_2^4. \label{soliton1}
\ee
where $\kappa^2= 2 \Lambda-m_2^2$.

 Under such conditions, $\psi_2$ describes a soliton confined within the interior of the black hole if
\bb
\kappa^2 >0.
\ee

\subsection{ Solution  $\psi_1^2 = -\frac{1}{\lambda_1}\left[ m_1^2 + \left(\frac{E_P}{2 M}\right)^2 g_1 g_2 \psi_2^2 \right]$}

This solution only contributes with redefinitions of the parameters that appear in the Lagrangian, which are essentially irrelevant. They are modifications of the type

\bb
2 \Lambda+ m_2^2 \to 2 \Lambda + m_2^2 + \frac{1}{2} m_1^2 \,g_1 \, g_2 \left(\frac{E_P}{2 M}\right)^2,
\ee
for the mass. For the coupling constant, the modification is

\bb
\lambda_2^2 \to \lambda_2^2 + {\cal O} \left[\left(\frac{E_P}{2 M}\right)^4\right],
\ee
which means the behavior of a stable and localized solution continues to persist inside the black hole.

In other words, solitons inside the black hole are intuitively consistent with general relativity, which states that nothing can escape from the interior unless the solutions become destabilized. Figure 1, shows a qualitative plot for higher powers in 
$\psi_2$ where the deformation for small values of $\gamma < 0$ is evident.
\subsection{A Philosophical Digression}

The Lagrangian (14) describes a $\lambda \varphi^4$ theory where the parameter $\kappa$ can be considered an IR scale that regulates the behavior of the theory at low energies (or large distances). If we consider the regime 
\[\Lambda \nsim m_2^2,\]
 then the cosmological constant ($\sqrt{|\Lambda|}$) is a natural scale for our problem.  Unlike $\Lambda_{\text{QCD}}$, the cosmological constant in this context is much more intriguing as it is an emergent scale that appears directly in the Lagrangian.
 
 The identification of this IR scale means that \footnote{As far as I know, the idea that the cosmological constant can be interpreted as an infrared cutoff first appeared in \cite{jurek}.}
 \bb
\fbox{$E_P = \sqrt{|\Lambda|}$}
\ee
 
 Thus, the expansion parameter of the effective theory we are considering is $\sqrt{|\Lambda|}/M \ll 1$, where $M$ is the UV cutoff associated with the mass of the $\chi$ field.
 
\section{ The correspondence between black holes and QCD}

To establish the analogy between QCD and black holes, we first note that the instantons arising from (1) are equivalent to the self-dual solutions of BPST, which are topological and are strongly dominant compared to the fermionic term \( \psi^\dagger (i \partial_4 + g A_4^a T^a - m) \psi \) (this is known as instanton dominance). This dominance allows us to identify the Lagrangian (\ref{ac2}) with (\ref{soliton1}) and hence draw the analogy that quarks in a hadron resemble matter inside a black hole described using the Kantowski-Sachs metric.

Instanton dominance does not mean that fermionic matter does not contribute; it states that fermions help modify the effective Lagrangian parameters while preserving their topological character.

Let us write (23) as 
\bb
{\cal L}_{eff} = \frac{1}{2} \dot{\psi}_2^2 +\frac{\lambda^2_2}{4} \left(\psi_2^2 - \frac{2 \kappa}{\lambda_2}\right)^2 +\cdots, \label{soli2}
\ee
 where $\cdots$ is a constant appropriately chosen to cancel the constant in (\ref{soli2}).

And redefining $\psi_2 = \frac{2 \kappa}{\lambda_2} \bar{\psi}_2$, the action becomes:
\bb
S= \frac{1}{\lambda^2_2} \int d{\bar t}\left[ \frac{1}{2} {\bar \psi}_2^2 + \frac{1}{4}\left({\bar \psi}_2^2-1\right)^2\right], \label{bh012}
\ee
which is precisely the correspondence between the black hole and QCD. This correspondence establishes that QCD confines because it has a causal horizon, just like a black hole.

It is remarkable to note that, by comparing (\ref{ac2}) with (\ref{bh012}), it is found that the description of QCD and black holes becomes equivalent if 
\[
\lambda_2^2 = \frac{g^2}{12 \pi^2},
\]
implying that 
\bb
\fbox{$\frac{|\Lambda|}{M^2} = \frac{1}{3\pi} \left(\frac{g}{g_2}\right)$}.
\ee

This is a very remarkable relation because it connects the cosmological constant with the Yang-Mills charge $g$, providing the link we were seeking to relate black holes with QCD. 
The correspondence rule is summarized in the table:
\[
\begin{array}{|c|c|}
\hline
\textbf{black Hole} & \textbf{QCD} \\ \hline
\text{horizon radius} \mapsto & \Lambda_{\text{QCD}}^{-1} \\ \hline
\text{black hole mass} \mapsto & \text{hadron mass} \\ \hline
\sqrt{\Lambda} \, (\text{cosmological constant}) \mapsto & \Lambda_{\text{QCD}} \\ \hline
\end{array}
\]

The effective Lagrangian for black holes and QCD is the same under the identifications outlined in Table 1. Consequently, if a black hole is confined due to the horizon acting as a causal barrier, QCD is also confined for the same underlying reasons. In the context of QCD, the black hole is a theoretical artifact to model the confinement phenomenon naturally, a hadron is not a black hole.

\section{Higher order corrections, deconfinement, and the black hole evaporation?}

Now, we will address the effect of higher-order corrections on effective action and whether these can overcome confinement. This is a non-trivial issue that requires careful analysis 
because resolving this problem would not only help to understand the meaning of deconfinement but also—since black holes play a fundamental role as a computational tool—provide insight 
into the phenomenon of radiation emission from the interior of black holes from a different perspective.

\subsection{Higher-order corrections to the effective action of black holes}

In the previous section, we have shown that if the first quantum corrections to the effective action describing the interior of a black hole are included, a soliton is formed, 
and the action is topological and stable. This last statement is because we have shown (\ref{ac2}) and (\ref{bh012}) are equivalent as a consequence of the self-duality condition.

Once this fact is established, the question is: can this action be destabilized or not by quantum corrections?

Before answering these questions, calculate these terms and start by minimizing the potential.
\eqb
V(\psi_1,\psi_2) &=& - 2 \Lambda \psi_2^2 +\frac{1}{2} G(0) \left(g_1 \psi_1^2 + g_2  \psi_2^2\right) - \frac{\gamma^4}{4} G^2(0) \left(g_1 \psi_1^2 + g_2  \psi_2^2\right)^2 \nonumber
\\ &-&
\frac{\gamma^8}{6} G^3(0) \left(g_1 \psi_1^2 + g_2  \psi_2^2\right)^3,
\eqf 
deriving respect to $\psi_1$ we obtain 
\bb
\frac{d V}{d \psi_1} = g_1G(o) \psi_1 \left( 1- \frac{1}{E_P M} y -\frac{1}{4 E_P^2 M^2}y^2\right)=0,
\ee
where we have defined $y= g_1\psi_1^2+ g_2 \psi_2^2$ and the solutions are 
\bb
\psi_1=0, ~~~~~~\text{and}~~~~\psi_1^2 = E_P M \delta_{\pm} - g_1 \psi_2,
\ee
where $\delta_\pm$ a real numerical coefficient.

In this way, the effective Lagrangian takes the form
\bb
{\cal L}_{eff} = \frac{1}{2} {\dot \psi_2}^2 - \frac{1}{2}\kappa~ \psi_2^2 + \frac{1}{4} \lambda_2^2 \psi_2^4 - \frac{1}{6}\gamma \psi_2^6+\cdots, \label{soliton12}
\ee
with $\gamma<0$.

Since $\gamma\, \psi_2^6$ is a perturbation added to a topological soliton, we might expect the soliton to experience some form of destabilization.  A very illustrative example is deconfinement, where the quark-gluon plasma forms a signal that exists only for short durations but leaves measurable traces in experiments.

The perturbation induces oscillations in the minimum energy state over short periods, which can eventually leave traces. These perturbations can escape the ``topological protection'' and manifest as indirect signals or ``flashes'' and that, in the specific case of deconfinement, emerges, for example, as jet quenching ({\it i.e.} high-energy quarks and gluons traversing a medium like the quark-gluon plasma lose energy, leaving traces of their interaction with the deconfined state) \cite{cao}.
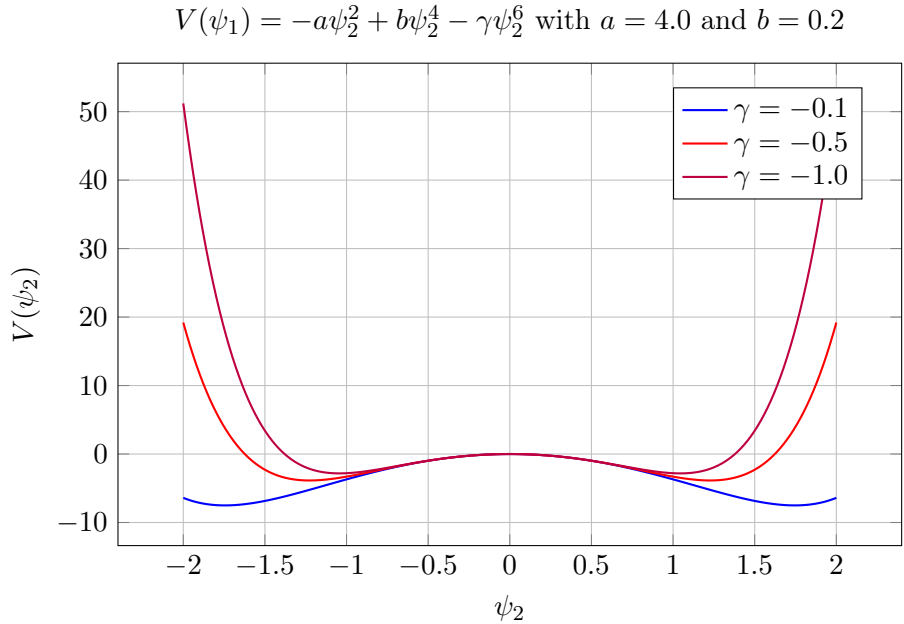
\begin{figure}
\begin{tikzpicture}
    \begin{axis}[
        width=12cm, height=8cm,
        domain=-2:2,
        samples=200,
        xlabel={$\psi_2$},
        ylabel={$V(\psi_2)$},
        title={$V(\psi_1) = - a \psi_2^2 + b \psi_2^4 - \gamma \psi_2^6$ with $a = 4.0$ and $b = 0.2$},
        legend style={at={(0.95,0.95)}, anchor=north east},
        grid=major,
    ]
        \addplot[color=blue, thick] expression{-4.0*x^2 + 0.2*x^4 - (-0.1)*x^6};
        \addlegendentry{$\gamma = -0.1$}
        
        \addplot[color=red, thick] expression{-4.0*x^2 + 0.2*x^4 - (-0.5)*x^6};
        \addlegendentry{$\gamma = -0.5$}
        
        \addplot[color=purple, thick] expression{-4.0*x^2 + 0.2*x^4 - (-1.0)*x^6};
        \addlegendentry{$\gamma = -1.0$}
        
    \end{axis}
\end{tikzpicture}
\caption{This caption provides a qualitative description of the behavior of the potential for several values of $\gamma$. Note that the potential spreads out as $|\gamma|$ decreases, and the soliton can leave the region in which it is confined. }
    \label{fig:potential_plot}
\end{figure}

We outline how this ``destabilization'' occurs; suppose we ``turn on'' $\gamma$, and it begins to grow slowly, causing the minima of the potential to shift toward the origin. On the other hand, if $\gamma$ continues to grow, the term $-\gamma\, \psi_2^6$ introduces a global instability, i.e., $V(\psi_2) \to -\infty$ as $\psi_2 \to \infty$.

\begin{figure}[h!]
    \centering
    \begin{tikzpicture}
        \begin{axis}[
            width=12cm, height=8cm,
            domain=-10:10,
            samples=200,
            xlabel={$\psi_2$},
            ylabel={$\mathcal{L}_{\text{eff}}$},
            title={Effective Lagrangian as a Function of $\psi_2$ with $\psi_2 \to \infty$},
            ymin=-1000, ymax=1000,
            legend style={at={(0.95,0.95)}, anchor=north east},
            grid=major,
        ]
            \addplot[color=blue, thick] expression{(2 * 1.0 + 1.0^2) * x^2 + (0.1 / 4) * x^4 + (-0.1 / 6) * x^6};
            \addlegendentry{Effective Lagrangian with large $\gamma<0$}
        \end{axis}
    \end{tikzpicture}
    \caption{Effective Lagrangian $\mathcal{L}_{\text{eff}}$ as a function of $\psi_2$ with large negative gamma, illustrating the divergence as $\psi_2 \to \infty$.}
    \label{fig:effective_lagrangian_divergence}
\end{figure}
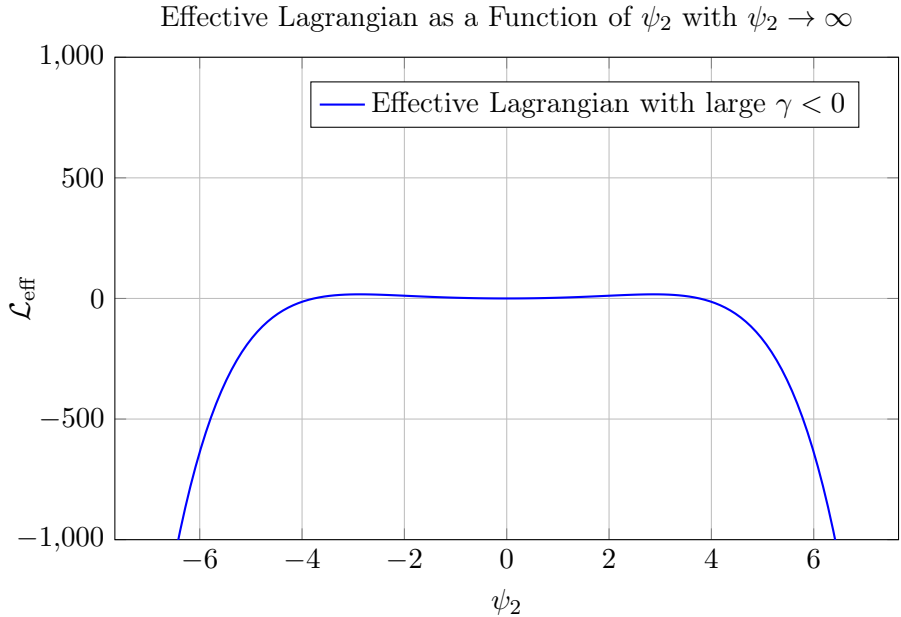

This ``evaporation'' occurs when $|\gamma|$ decreases; in that case, the confinement boundaries become blurred, and localization is lost. What is the probability of escape? To answer this question, we would have to solve the equation of motion for $\psi_2$ and calculate $e^{-S[\psi_1]}$, which is analytically impossible. Figure 2 roughly shows this behavior, although understanding, of course, that the disturbances are occurring on a topologically protected system.

However, if $\gamma$ grows very slowly, destabilization will occur, and eventually, it will escape from the interior with a minimal probability (but non-zero). However, it is remarkable that this approach provides a completely different perspective on the information paradox and, on the other hand, also offers an alternative to Hawking radiation.
\subsection{ Deconfinement of  Black Holes}

In this subsection, we present a physical framework to visualize how a black hole could undergo deconfinement. Suppose that, for some reason, a black hole experiences a massive ``shake'',  causing it to vibrate and partially eject matter from a certain region. However, due to the black hole's immense gravitational field, most of the ejected matter is reabsorbed. Nevertheless, during this event, some matter may achieve a velocity greater than the escape velocity in the region where the ``extreme oscillations'' occur. While this escaping matter is an infinitesimally small amount, it constitutes part of the information released by the black hole.

One can compare this process to Hawking radiation, a quantum mechanism through which black holes emit thermal radiation. In contrast, the dynamic escape of matter during perturbative events, such as large extreme oscillations offers a complementary mechanism for releasing specific information about the black hole's interior, potentially addressing the information paradox.

Just as quark deconfinement leaves signals such as an overabundance of hadronic particles, jet suppression, and other phenomena, we might consider that large oscillations could produce gravitational waves as potential signals.

\section{discussion}

There is a crucial aspect that we highlight: the effective Lagrangian provides instanton solutions and is asymptotically free, as demonstrated by the $\beta$-function \[ \beta (M) = M\frac{\partial \lambda_2^2}{\partial M} = - \frac{1}{2} \lambda_2^2 (M) \].  

Remarkably, according to our description, the physical theory inside a black hole is asymptotically free (possibly due to the Carroll limit). This implies that quantum treatment within the black hole is simplified. However, this result should not surprise us because, as already discussed, the black hole described as an effective theory is formally equivalent to QCD with instanton dominance.

From this research perspective, the interior of a black hole can be described by a scalar field theory in a practically flat space within the Carroll limit. The absence of curvature effects arises from selecting well-suited variables to avoid mathematical difficulties. Since the space is flat, solitons can be destabilized by adding quantum corrections, which, although small, remain finite (a Carroll quantum theory does not require renormalization). The higher-order terms asymptotically exhibit the form \( \psi_2^6 \) and beyond, which, while extremely small, ultimately lead to black hole evaporation. 

In summary, our main results are:
\begin{enumerate}
\item We have constructed an effective theory for the interior of black holes, demonstrating that the interior can be viewed as an effective theory describing solitons.\item This theory is mappable to Yang-Mills theory, and consequently, if a black hole confines due to the presence of causal horizons, then QCD confines for the same reasons.
\end{enumerate}

 \section{Acknowledgments}
 The author thanks M. B. Paranjape and R. B. Mackenzie for their hospitality during his visit to Montreal. He also thanks J. L. Cortés, J. Lopez-Sarrion, F. Méndez, and I. Obreque for the valuable discussions. This research was supported by Fondecyt 1221463 and Dicyt (USACH).


\begin{thebibliography}{99}
\bibitem{levy} J. -M. Lev\'y-Leblond, Annales de l'I. H. Poincar\'e section A, tome 3, no 1 (1965), p. 1-12.
\bibitem{rev1} See for example L.~Donnay and C.~Marteau,
Class. Quant. Grav. \textbf{36} (2019) no.16, 165002
doi:10.1088/1361-6382/ab2fd5
[arXiv:1903.09654 [hep-th]]; M.~Henneaux and P.~Salgado-Rebolledo,
JHEP \textbf{11} (2021), 180
doi:10.1007/JHEP11(2021)180
[arXiv:2109.06708 [hep-th]]; E.~A.~Bergshoeff, A.~Campoleoni, A.~Fontanella, L.~Mele and J.~Rosseel,
PoS \textbf{CORFU2023} (2024), 235
doi:10.22323/1.463.0235; A.~Bagchi, S.~Banerjee, R.~Basu and S.~Dutta,
Phys. Rev. Lett. \textbf{128} (2022) no.24, 241601
doi:10.1103/PhysRevLett.128.241601
[arXiv:2202.08438 [hep-th]] and references therein.
\bibitem{polyakov} A.~A.~Belavin, A.~M.~Polyakov, A.~S.~Schwartz and Y.~S.~Tyupkin,
Phys. Lett. B \textbf{59} (1975), 85-87
doi:10.1016/0370-2693(75)90163-X.

\bibitem{review1} G.~'t Hooft,
Nucl. Phys. B \textbf{190} (1981), 455-478
doi:10.1016/0550-3213(81)90442-9; L.~Susskind,
Phys. Rev. D \textbf{20} (1979), 2610-2618
doi:10.1103/PhysRevD.20.2610; G.~'t Hooft,
Nucl. Phys. B \textbf{138} (1978), 1-25
doi:10.1016/0550-3213(78)90153-0
\bibitem{KS} R.~Kantowski and R.~K.~Sachs,
J. Math. Phys. \textbf{7} (1966), 443
doi:10.1063/1.1704952; see also, 

\bibitem{brehme}  R.~W.~Brehme,
Am. J. Phys. \textbf{45} (1977), 423
doi:10.1119/1.10829.
\bibitem{klauder} J. R. Klauder, Commun. Math. Phys. \textbf{18} (1970), 307-318

\bibitem{justo} J.~Gamboa and J.~Lopez-Sarrion,
Int. J. Mod. Phys. A \textbf{36} (2021) no.13, 13
doi:10.1142/S0217751X21500743
[arXiv:2009.05852 [hep-th]].

\bibitem{calvo} M.~Calvo,
Phys. Rev. D \textbf{15} (1977), 1733-1735
doi:10.1103/PhysRevD.15.1733.

\bibitem{coleman} S.~R.~Coleman,
``The Uses of Instantons,''
Subnucl. Ser. \textbf{15} (1979), 805
HUTP-78-A004.


\bibitem{thooft1} G.~'t Hooft,
Phys. Rev. D \textbf{14} (1976), 3432-3450
[erratum: Phys. Rev. D \textbf{18} (1978), 2199]
doi:10.1103/PhysRevD.14.3432.

\bibitem{diakonov} D.~Diakonov,
Prog. Part. Nucl. Phys. \textbf{51} (2003), 173-222
doi:10.1016/S0146-6410(03)90014-7
[arXiv:hep-ph/0212026 [hep-ph]]; A.~Actor,
Rev. Mod. Phys. \textbf{51} (1979), 461
doi:10.1103/RevModPhys.51.461.
\bibitem{maldacena} O.~Aharony, S.~S.~Gubser, J.~M.~Maldacena, H.~Ooguri and Y.~Oz,
Phys. Rept. \textbf{323} (2000), 183-386
doi:10.1016/S0370-1573(99)00083-6
[arXiv:hep-th/9905111 [hep-th]].

\bibitem{vile} G.~Fanaras and A.~Vilenkin,
JCAP \textbf{03} (2022) no.03, 056
doi:10.1088/1475-7516/2022/03/056
[arXiv:2112.00919 [gr-qc]]; J.~J.~Halliwell and J.~Louko,
Phys. Rev. D \textbf{42} (1990), 3997-4031
doi:10.1103/PhysRevD.42.3997; D.~W.~Chiou,
Phys. Rev. D \textbf{78} (2008), 044019
doi:10.1103/PhysRevD.78.044019
[arXiv:0803.3659 [gr-qc]]; E.~Weber,
J. Math. Phys. \textbf{25} (1984), 3279
doi:10.1063/1.526076.
\bibitem{jurek} J.Kowalski-Glikman and L.Smolin,
Phys. Rev. D \textbf{70} (2004), 065020
doi:10.1103/PhysRevD.70.065020
[arXiv:hep-th/0406276 [hep-th]].
\bibitem{cao} S.~Cao and X.~N.~Wang,
Rept. Prog. Phys. \textbf{84} (2021) no.2, 024301
doi:10.1088/1361-6633/abc22b
[arXiv:2002.04028 [hep-ph]].

  \end{thebibliography}
\end{document}